\pdfoutput=1

\documentclass[sigconf,nonacm]{acmart}

\usepackage{graphicx}
\usepackage{hyperref}
\usepackage{color}
\usepackage{booktabs}
\usepackage{amsmath}

\AtBeginDocument{%
  }

\begin{document}
\title[Ranking Narrative Query Graphs for Biomedical Document Retrieval (Technical Report)]{Ranking Narrative Query Graphs for \\ Biomedical Document Retrieval (Technical Report)}

\author{Hermann Kroll}
\email{krollh@acm.org}
\orcid{0000-0001-9887-9276}
\affiliation{%
  \institution{Institute for Information Systems, \\ TU Braunschweig}
  \country{Germany}
}

\author{Pascal Sackhoff}
\email{p.sackhoff@tu-bs.de}
\orcid{0009-0005-3095-9794}
\affiliation{%
  \institution{Institute for Information Systems, \\ TU Braunschweig}
  \country{Germany}
}

\author{Timo Breuer}
\email{timo.breuer@th-koeln.de}
\affiliation{%
  \institution{TH Köln (University of Applied Sciences), Germany}
  \country{Germany}
}

\author{Ralf Schenkel}
\email{schenkel@uni-trier.de}
\orcid{0000-0001-5379-5191}
\affiliation{%
  \institution{Universität Trier}
  \country{Germany}
}

\author{Wolf-Tilo Balke}
\email{balke@ifis.cs.tu-bs.de}
\orcid{0000-0002-5443-1215}
\affiliation{%
  \institution{Institute for Information Systems, \\\ TU Braunschweig}
  \country{Germany}
}

\renewcommand{\shortauthors}{Kroll et al.}

\begin{abstract}
Keyword-based searches are today's standard in digital libraries. Yet, complex retrieval scenarios like in scientific knowledge bases, need more sophisticated access paths. Although each document somewhat contributes to a domain's body of knowledge, the exact structure between keywords, i.e., their possible relationships, and the contexts spanned within each single document will be crucial for effective retrieval. Following this logic, individual documents can be seen as small-scale knowledge graphs on which graph queries can provide focused document retrieval. We implemented a full-fledged graph-based discovery system for the biomedical domain  and demonstrated its benefits in the past. Unfortunately, graph-based retrieval methods generally follow an 'exact match' paradigm, which severely hampers search efficiency, since exact match results are hard to rank by relevance. This paper extends our existing discovery system and contributes effective graph-based unsupervised ranking methods, a new query relaxation paradigm, and ontological rewriting. These extensions improve the system further so that users can retrieve results with higher precision and higher recall due to partial matching and ontological rewriting.
\end{abstract}

\ccsdesc[500]{Information systems~Information retrieval}
\ccsdesc[300]{Information systems~Document representation}
\ccsdesc[100]{Information systems~Digital libraries and archives}

\keywords{Graph-based Ranking, Document Retrieval, Biomedical Retrieval, Information Retrieval, Digital Libraries}

\maketitle

\section{Introduction}
Digital libraries usually implement document retrieval through simple-to-use keyword-based access paths. 
However, in complex retrieval scenarios like for scientific documents, the use of learning architectures to learn the relevance between a user's query and individual textual documents can severely boost retrieval performance~\cite{albarede2022gatforpassageretrieval,hui2017pacrr,lu2013deepmatch,mohan2018deeplearningforbioir}. 
In a nutshell, such systems use a first stage for initial retrieval and then apply strategies like \textit{neural re-ranking} or \textit{learning-to-rank} to estimate the relevance of documents, e.g.,  ~\cite{hui2017pacrr,lu2013deepmatch,mohan2018deeplearningforbioir,DBLP:journals/corr/abs-1904-08375,DBLP:journals/corr/abs-2101-05667,DBLP:journals/corr/abs-2212-03533}.
Although these approaches proved to be very effective on different benchmarks, they come with two major limitations: 
First, a large quantity of training data needs to be provided to learn how the documents' relevance relates to individual queries. 
Second, applying deep learning in large-scale scenarios is quite costly: acquiring training data, training time, hardware, etc.

Building on the success of large knowledge graphs, a viable alternative is to adapt the graph-based retrieval paradigm to IR-style document retrieval. 
In the past, we proposed so-called narrative query graphs, see~\cite{Kroll2023IJDL,kroll2022nia} and \url{www.narrative.pubpharm.de} for an implementation in the field of bio-medicine. 
Here, users can represent information needs as directed \textit{edge-labeled graphs}. 
This intuitive kind of querying means simply stating relevant concepts and their interactions and can be supported by suitable user interfaces~\cite{Kroll2023IJDL}. 
The resulting query graph representation is then matched against a large set of focused document graphs, each individually extracted from some document in the digital library. 
In contrast to traditional knowledge graph querying, where all extracted information is integrated into one big knowledge graph, document-centered graph query processing ensures the validity of results through context-compatible information fusion~\cite{kroll2020tpdl}. 
That means that narrative graph queries are only answered in strict document contexts, i.e., by fusing statements mentioned within the scope of a single document. 
However, the graph-based retrieval approach also suffers from a severe limitation: queries are isomorphically matched against document graphs, i.e., all correct answers show the same level of relevance (exact matches).

\textit{So, how can graph-based document representations be effectively ranked for document retrieval purposes?}
This paper extends our existing system by introducing novel ranking strategies that 1) intelligently exploit the structure of document graph representations and 2) effectively increase the retrieval recall through a relaxed query matching paradigm (Partial Matches) and ontological query rewriting. 
Moreover, our methods do not rely on supervision and can thus be deployed without requiring costly training data. 
To demonstrate our approach's effectiveness, we evaluated it on five biomedical benchmarks:
The TREC Precision Medicine Series (2017-2020)~\cite{DBLP:conf/trec/RobertsDVHBLP17,DBLP:conf/trec/RobertsDVHBL18,DBLP:conf/trec/RobertsDVHBLPM19,DBLP:conf/trec/RobertsDVBH20} includes concept-centric queries that show advantages of our method.
For instance, the 2020 version of the benchmark series asks for precise gene-disease-treatment combinations.
To also show the limitations of our approach, we selected the TREC-COVID 2020~\cite{DBLP:journals/jbi/RobertsABDLSVWH21} benchmark, which includes rather generic query formulations formulated by non-domain experts, i.e., precise concepts were missing or information needs like \textit{closing of schools} are not reflected in the system's vocabulary. 
We compare our methods to BM25 ranking.

While the systems implementation's query processing is currently restricted to the biomedical domain and concept-centric queries, this paper is the first to investigate a full-fledged ranked graph-based document retrieval system from an information retrieval perspective. 
In addition to our graph-based discovery system's benefits like structured literature overviews, see~\cite{Kroll2023IJDL} for a comprehensive overview, this paper proposes graph-based ranking methods and query relaxation strategies to improve such a system significantly, in terms of precision and recall. 
For that, we do not claim that we outperform existing information retrieval techniques, especially learned systems.
Instead, we extended our existing real digital library system by ranking methods and expansion that is desirable for users. 
We share our code, produced results and detailed topic-wise evaluation figures at GitHub\footnote{\url{https://github.com/HermannKroll/RankingNarrativeQueryGraphs}} and Software Hertiage\footnote{Software Heritage ID:\href{https://archive.softwareheritage.org/swh:1:dir:56036430260e2759be3ac9b72f6160fed361f503}{swh:1:dir:56036430260e2759be3ac9b72f6160fed361f503}}. Our graph-based discovery system is available at\footnote{\url{https://narrative.pubpharm.de}}.

\section{Related Work}
\label{sec:related_work}
\paragraph{Information Retrieval.}
Information retrieval is the task of finding relevant information concerning some user's information need~\cite{DBLP:books/daglib/0021593}.
In a digital library, one way to support users is to implement Boolean term-based retrieval, i.e., a user states a set of terms, and a document is considered relevant if the terms are contained.
In addition to that, the relevance of some documents can be estimated to create a ranked list of results~\cite{DBLP:books/daglib/0021593}, e.g., via vector space models, tf-idf-based metrics or BM25 rankings. 
BM25 is a common method that is rather cheap to implement and effective. 
It builds upon tf-idf plus document length weighting.
Query rewriting and expansion allow more sophisticated retrieval, i.e., by rewriting the query to a precise one or expanding it via synonyms to retrieve more results.

Today, neural methods are used to learn the relevance between a user's query and individual textual documents.
Learning suitable relevancy can indeed severely boost retrieval performance~\cite{albarede2022gatforpassageretrieval,hui2017pacrr,lu2013deepmatch,mohan2018deeplearningforbioir}. 
Usually, such systems use a first stage for initial retrieval and then apply strategies like \textit{neural re-ranking} or \textit{learning-to-rank} to estimate the relevance of documents, e.g.,  ~\cite{hui2017pacrr,lu2013deepmatch,mohan2018deeplearningforbioir,DBLP:journals/corr/abs-1904-08375,DBLP:journals/corr/abs-2101-05667,DBLP:journals/corr/abs-2212-03533}.
However, deep learning systems usually have two major limitations: On the one hand, they require sufficient and high-quality training data to learn a suitable notion of relevance.
Second, these systems are hard to explain: Why is document A more relevant than document B, or why is A even considered relevant?
Explaining is the objective of research, but often challenging to implement in a digital library.

Instead of relying on neural methods, our approach extends a graph-based discovery system that explains document matches to the user. 
In brief, users formulate some query patterns, and a matching document must contain precisely that pattern.
Beyond that, the system can show the original text span from which the pattern has been extracted so that users can quickly see whether the document is relevant to them.

\paragraph{Biomedical Information Retrieval.}
Generally speaking, traditional approaches especially for the TREC Precision Medicine benchmark series, often build on BM25~\cite{DBLP:conf/trec/NohK17,DBLP:conf/trec/ZhengLHX18,DBLP:conf/trec/NunzioMA19,DBLP:conf/trec/NguyenKFAPW17,DBLP:conf/trec/AgostiN018,DBLP:conf/trec/RybinskiKP19,DBLP:conf/trec/RobertsDVBH20}.
Yet, they often integrate additional domain-specific knowledge like controlled vocabularies, identified concepts, and special rules.
However, those rules are usually hand-crafted for certain information, e.g., how to expand concrete demographic information (38-aged-male) with relevant target groups (males between 30 and 40) or search for similar genes. 
Recently, two notable graph-based systems have been designed for the biomedical domain: Ebeid and Pierce~\cite{ebeid2021medgraph} construct a knowledge graph involving metadata and entity mentions in texts for biomedical information retrieval. The entity information is then used to expand a user's query by relevant entities. 
GRAPHENE~\cite{zhao2019graphenebiomedicalretrieval} is a graph-augmented document representation learning approach for query expansion. 
A learning-to-rank architecture is developed to improve the retrieval of biomedical documents.
In contrast to these approaches, our methods are unsupervised and are based on extracted statements from text, i.e., on pure document graphs.

\paragraph{Graph-based Retrieval.}
Graph-based retrieval has been shown to improve retrieval for a variety of applications~\cite{dietz2018kgfortextretrieval,zhao2019graphenebiomedicalretrieval,ebeid2021medgraph,kadry2017openreforretrieval,albarede2022gatforpassageretrieval}. 
Dietz et al. examined how knowledge graphs could be used for text-centric information retrieval~\cite{dietz2018kgfortextretrieval} discussing entities, graph structures, and relations. 
Features like known metadata or entity text annotations even allowed some systems to learn more sophisticated relevance scores~\cite{zhao2019graphenebiomedicalretrieval}. 
Another work discussed how open relation extraction might accelerate passage retrieval for given entities in queries~\cite{kadry2017openreforretrieval}.
Although these works are strongly connected to ours, they rather suggest features and first evaluations instead of proposing ready-to-use methods. 
Instead, we extended a real-world digital library system, here at the showcase of our full-fledged graph-based discovery system, by integrating a suitable graph-based ranking method and expansion paradigms.

\paragraph{Narrative Query Graphs.} 
Our and PubPharm's (German specialized information service for pharmacy) graph-based retrieval service, called the Narrative Service~\cite{Kroll2023IJDL}, currently features approx. 37 million publications from the National Library of Medicine's Medline collection and 70k COVID-19 pre-prints from ZB MED's preVIEW service~\cite{langnickel2021covid}. 
Users are enabled to intuitively formulate their information needs as graph patterns, i.e., conjunctions of triple-like statements (concept, interaction, concept). 
Correct answers to the query are all documents that contain all search statements. 
For the query processing, document texts are transformed into a graph representation by linking terms against biomedical concepts and extracting their interactions through the PathIE method~\cite{DBLP:conf/jcdl/KrollPB21}. 
Concepts were identified by deriving annotations from the PubTator service~\cite{wei2013pubtator,wei2019pubtatorcentral} and performing a dictionary-based concept linking through vocabularies derived from ChEBML~\cite{mendez2018chembl}, Wikidata~\cite{vrandecic2014wikidata} and the Medical Subject Headings. 
PathIE extracted a statement between two concepts as follows:
Given a sentence with two detected concepts, the dependency parse of the sentence (basically the grammatical structure) was computed using the Stanford CoreNLP toolkit~\cite{DBLP:conf/acl/ManningSBFBM14}. 
A triple-statement was extracted if concepts were connected through a verb phrase or particular keywords from a pre-defined list (e.g., treatment, therapy, inhibition). 
Verb phrases were canonicalized by a hand-crafted relation vocabulary (relations plus synonyms) and a word embedding to compute the similarity of verbs. 
Overall, due to the lack of supervision, the service comes with a moderate linking quality between 56.0\% and 77.6\% F1 (four benchmarks) and an extraction quality of 39.1\% F1 (two benchmarks) as analyzed in~\cite{DBLP:conf/jcdl/KrollPB21}. 
Beyond specific interactions such as \textit{treats} or \textit{inhibits}, a sentence-based extraction method extracts general \textit{association} statements if two concepts were mentioned within the same sentence. 
The confidence of such extractions within a sentence is defined as the distance (character count) between both concepts divided by the sentence length.

\section{Graph-based Retrieval and Ranking}
In this section, we introduce basic notations and our graph-based ranking methods. 
We described the details of the graph-based retrieval service we are going to extend in Sect.~\ref{sec:related_work}. 
The retrieval is based on the interactions between biomedical concepts. 
Formally, $\mathcal{C}$ is the set of known \textbf{concepts} (e.g., Metformin, Diabetes), and $\Sigma$ is the set of known \textbf{interactions} (e.g., associated, treats, inhibits).
A \textbf{statement} is an triple ($c_1$, $p_1$, $c_2$) with $c1, c2 \in \mathcal{C}$ and $p_1 \in \Sigma$ .
Each \textbf{document} is represented by its so-called document graph, which is harvested from its title and abstract. 
A \textbf{document graph} $graph(d)$ is a directed, edge-labeled graph extracted from the corresponding document $d$.
Please note that an edge could be \textit{extracted} from several sentences of $d$.
Each of these extractions comes with a confidence score, e.g.,  the applied extraction method PathIE defines \textbf{confidence} over the distance between two concepts in the grammatical structure of a sentence (concepts with a close grammatical connection receive higher confidence). 

\begin{figure*}[t]
    \centering
    \includegraphics[width=\textwidth, trim={0cm 11cm 0cm 0cm}, clip]{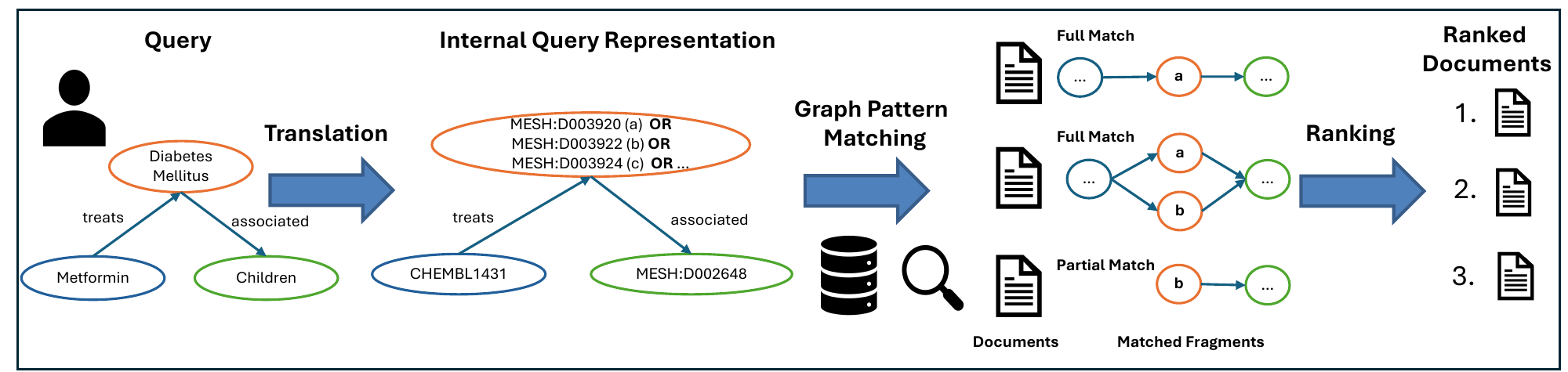}
    \caption{Systematic overview: Users formulate their information needs as graph patterns between concepts. Queries are translated and matched against document graphs. Matches are documents that match the query completely (full match) or partially (partial match). The matched documents are then ranked based on their graphs.}
     \Description{Systematic overview: Users formulate their information needs as graph patterns between concepts. Queries are translated and matched against document graphs. Matches are documents that match the query completely (full match) or partially (partial match). The matched documents are then ranked based on their graphs.}
    \label{fig:graphbasedranking}
\end{figure*}

The Narrative Service allows users to formulate their information needs as narrative query graphs. 
Here, we summarize the key aspects of the formal mechanism, originally published in~\cite{Kroll2023IJDL}.
A \textit{narrative query} consists of a set of fact patterns.
A \textbf{fact pattern} is a triple $(s, p, o)$. 
The subject $s$ and object $o$ are either concepts from $\mathcal{C}$ or variables from a set $\mathcal{V}$. 
The fact patterns are understood as being logically connected via an AND expression. 
If the narrative query graph does not ask for variables, matches to the query are documents that contain all searched $(s, p, o)$ triples in their document graph.
If a narrative query graph contains variables, a document $d$ matches the query if 1) the function $\mu_d: \mathcal{V} \rightarrow \mathcal{C}$ substitutes the query's variables with concrete concepts from $\mathcal{C}$ and 2) the resulting, substituted graph with concrete concepts is supported by the document graph of $d$, i.e., the substituted query pattern is part of $d$'s graph. 

For this paper, we focus on ranking narrative query graphs without variables. 
Queries with variables are answered by aggregated substitution lists, i.e., a list of documents that share an equal substitution. 
Our existing system ranks those lists by the number of documents that share the same substitutions that users accept. 
User studies verified that such a ranking is sufficient for our users~\cite{Kroll2023IJDL}.

\subsection{Query Translation} 
At the moment, our discovery system comes with a query builder:
Users enter a desired concept as a human-readable label instead of artificial concept identifiers. 
An autocompletion function assists users in knowing which terms are known by the system.
A selection box visualizes different interactions (e.g., treats, inhibits) for the users.
If a fact pattern is completed this way, users can start their search or enter another pattern. 
The system then translates the user input into an internal query representation. 
However, mapping an entered label to precise concept IDs is not trivial. 
Consider, for example, a concept search like \textit{diabetes}. 
In that case, users typically would expect that all texts that contain the word \textit{diabetes} are found. 
In the current implementation, the system thus translates the user input by mapping the entered term to all concepts with a synonym that starts with the searched term.
For the term \textit{diabetes}, the system would conduct a search for the gene \textit{diabetes} and the diseases \textit{diabetes mellitus}, \textit{diabetes mellitus type 1}, and \textit{diabetes mellitus type 2}.
In this way, \textit{diabetes} will be mapped to the previously introduced concepts but not to \textit{gestational diabetes} because it does not start with \textit{diabetes}.
\textit{Gestational diabetes} is a special subtype of \textit{diabetes mellitus}. 
The system automatically expands searches, i.e., if a search for a concept is conducted, concepts that are subclasses of that concept are included by default.

In this paper, we improve the query translation process in two ways: 
First, all concepts containing the searched term in one of the synonyms are now considered relevant.
The advantage is that it does not matter whether the user enters \textit{diabetes mellitus} or \textit{mellitus diabetes}.
We implement the strategy by using a relational database table that maps terms to concepts.
In the case of a \textit{diabetes} search, the table is queried by a SQL WHERE expression:  \textit{term LIKE \%diabetes\%}.
Suppose the user's entered term contains multiple terms. In that case, the terms are split by space and concatenated by  AND operations, e.g., \textit{diabetes mellitus} is translated into a SQL WHERE expression like \textit{term LIKE \%diabetes\% AND term LIKE \%mellitus\%}. 
This strategy ensures that all entered terms are contained, but the order is not essential, making the translation easier to use, and in some cases, more robust, e.g., it does not matter whether the user searches for \textit{diabetes mellitus} or \textit{mellitus diabetes}.
We created a trigram-based index to accelerate LIKE searches in the database.

Second, we introduce a so-called \textit{translation score} to represent how well a translated concept might represent the user's intended concept search. 
If the user input directly matches a synonym, it is considered a perfect concept translation.
The more the strings (user input and synonym of a concept) differ, the less accurate the translation is.
We used the Jaccard string similarity as one possible implementation strategy to measure the string distance, as it measures how many terms of the user input correspond to how many terms of some concept vocabulary entry.
Let $u$ be the user's input, and $c$ be a concept that contains $u$ in one of its synonyms $s$. 
The translation score is defined as $\textbf{translation\_score}(c, u) = \textit{jaccard\_similarity}(u, s)$.
We could have used different semantic measures, like vector space distances, to determine which concepts best match the user's input.
However, our graph-based system focuses on biomedical concepts; users are expected to search for those concepts.
That is why we decided that users must enter the  concept names, and a string-based similarity, such as Jaccard, is sufficient here.
Exploring alternatives could be worth further investigation.

In brief, the translation process could yield the following result:
A users enters a search like (\textit{Metformin}, \textit{treats}, \textit{Diabetes}).
The translated query is $qt = (c_{Metformin}, \textit{treats}, (c_{D} \lor c_{DM} \lor c_{DMT1} \lor c_{DMT2}))$, i.e., the query ask for a \textit{Metformin} concept and four possible alternatives for the \textit{diabetes} translation (the concepts $c_D$, $c_{DM}$, $c_{DMT1}$ and $c_{DMT2}$).

\subsection{GraphRank}
In the following, we propose our graph-based ranking method \textbf{GraphRank}. 
As stated above, a translated query is a conjunction of fact patterns, where 
each fact pattern might ask for alternative concepts for subjects and objects (disjunction). 
In the first step, we thus have to determine which part of a document graph matches the query.
Suppose a query like \textit{Metformin treats Diabetes Mellitus}.
In that case, the term \textit{Diabetes Mellitus} could be directly matched to the concept \textit{Diabetes Mellitus}.
However, it could also be matched to \textit{Diabetes Mellitus, Type 1} (as explained in our previous section).

Due to alternative concepts in each given expansion, a document graph might thus match with different graph parts (different subgraphs), e.g., one match might include \textit{Diabetes Mellitus type 1} and another \textit{type 2}. 
The function $matches(q, d)$ computes all distinct subgraph isomorphisms between the query $q$ and the document graph of $d$. 
Each subgraph isomorphism maps a part of the document graph to the query. 
We call that matching part \textbf{fragment}, i.e., given a query $q$, a document $d$, and a fragment $f \in matches(q, d)$. 
The function $\textit{edges}(f)$ returns edges of the fragment $f$, and $\textit{nodes}(f)$ returns the nodes of $f$.
In other words: A fragment is a subgraph of the document graph that matches the query. 
Please note that a document can have multiple fragments because different document graph concepts can match the same query.

\paragraph{Extraction confidence} 
Our discovery system transforms a document's text into a graph representation.
By doing that, information extraction methods are used that come with some confidence score, i.e., how sure the system was about the extraction.
The applied extraction method PathIE defines \textbf{confidence} over the distance between two concepts in the grammatical structure of a sentence (concepts with a close grammatical connection receive higher confidence). 
For \textit{associated}-statements, the confidence within a sentence is defined as the distance (character count) between both concepts divided by the sentence length.
Next, the same statement, e.g., Metformin treats Diabetes Mellitus, could be extracted from multiple sentences within a document.
Multiple extractions might thus support some document graph's edge.
We decided to use the maximum confidence score to represent the statement's final confidence score (it was the best extraction for that statement within $d$).
Hence, we define the confidence for an edge $e$ of a document $d$ as:

\begin{equation}
    \textit{edge\_conf}(e, d) = \max(\lbrace \textit{conf(s)} \mid s \in \textit{statements} (d) \land s \textit{ supports } e \rbrace)
\end{equation}

Next, we score a fragment $f$ to be only as strong as its weakest edge, i.e., only as good as the less confident statement extraction:

\begin{equation}
  \textit{confidence}(f, d) = \min(\lbrace \textit{edge\_conf}(e, d) \mid e \in \textit{edges}(f) \rbrace)
\end{equation}

\paragraph{tf-idf statement.}
Inspired by traditional information retrieval methods, we design a tf-idf strategy to score fragments. 
The idea is that some edges carry more information (are more relevant) than very general ones. 
However, maintaining an idf index for statements comes with two problems: 
First, the index can get quite large (quadratic growth with regard to the size of the concept vocabulary).
Second, statement extraction is limited to sentences and might be error-prone. Many connections might be lost during that step, affecting the idf score.
That is why we decided to approximate the tf-score for an edge by combining the tf-idf scores of its subject and object plus multiplying it with a predicate specificity, i.e., a score between 0 and 1 that determines how specific a predicate is.

We define tf for a concept $c$ within a document $d$ as $\textit{tf}(c, d) = \frac{\#(c, d)}{\#(c', d)}$ with $\#(c, d)$ being the number of occurrences of concept c within $d$ and $c'$ being the concept that has the maximum number of occurrences (for normalization).
Next, we define idf for a concept $c$ as $\textit{idf}(c) = \log \frac{|\mathcal{D}|}{|\{d \in \mathcal{D}\land c \in d\}|}$. 
A point to consider is whether idf is still required because we already search for biomedical concepts, in contrast to arbitrary terms like stopwords.
Some biomedical concepts like \textit{humans} appear in millions of documents, while others like specific genes may appear only in a few documents.
Suppose a query asks for an interaction between such a gene and humans. 
In that case, the gene should influence the overall ranking more, i.e., a document mentioning the gene more often is better than a document mentioning the human more often.
That is why we decided also to use idf in our approach.
$|\mathcal{D}|$ is the number of documents in our collection, and the denominator counts documents that include the concept $c$. 
With that, we can define the tf-idf score for an edge $e = (s, p, o)$ concerning a document $d$ as: 

\begin{equation}
    \textit{edge\_tfidf}(e, d) = (\textit{tf}(s, d) \cdot \textit{idf}(s) + \textit{tf}(o, d) \cdot \textit{idf}(o)) \cdot \textit{specificity}(p)
\end{equation}

A fragment $f$ of some document $d$ might include multiple edges. 
Our idea is that the minimum score represents the weakest link in the fragment, and we define the fragment's tf-idf score as the minimum edge score:

\begin{equation}
 \textit{min\_tfidf(f, d)} = min(\lbrace \textit{edge\_tfidf}(e, d) \mid e \in \textit{edges}(f)) \rbrace
\end{equation}

\paragraph{Coverage.} 
Article abstracts may often start with some background information in the corresponding field. 
Concept mentions within that background part might be less important than concepts mentioned across the whole abstract.
We therefore propose a method that considers how much \textit{coverage} a concept $c$ has within a document $d$:

\begin{equation}
    \textit{c\_coverage}(c, d) = \frac{\textit{last\_position}(c, d) - \textit{first\_position}(c, d)}{\textit{text\_length}(d)}
\end{equation}

The method calculates the difference between the concept's last mention and the first mention within the document and normalizes it by the document's text length.
Coverage approximates whether the concept is used from the beginning to the end or briefly mentioned somewhere as a side note.
The higher the coverage is, the more relevant a concept $c$ should be.
We then define the coverage of a fragment $f$ with regard to $d$ by its weakest \textit{covered} concept:

\begin{equation}
    \textit{coverage}(f, d) = \min(\lbrace \textit{c\_coverage}(c, d) \mid c \in \textit{nodes}(f) \rbrace)
\end{equation}

Please note, that coverage will only be a part of our final ranking method. 
We still consider the occurrences of some concepts through their tf-idf scores.
In other words: Although a document may receive a high coverage score because a concept appears in the beginning and in the end, the tf score can still be low because the concept only appears twice.

\paragraph{Relational similarity.}
The neighborhood of an edge $e$, i.e., incoming and outgoing edges to the edge's subject or object, could be relevant to estimate the relevancy of the edge $e$. 
The function $h$ takes an edge $e = (s, p, o)$ and returns the neighbour edges: $h(e, d) = \lbrace (s', p', o') \mid (s', p', o') \in graph(d) \land (s' = s \oplus o' = o)  \rbrace$, i.e., edges that are incoming or outgoing to $s$ and $o$, but no edges between $s$ and $o$. 
We compute the relational similarity of an edge $e$ by averaging all edge neighbors' tf-idf, coverage, and confidence scores (given by $\textit{edge\_score}$).
The coverage of an edge is defined by the minimum coverage of one of its concepts (similar to our coverage definition for fragments). 
With that, we define the relational similarity of a fragment $f$:

\begin{equation}
    \textit{relational\_similarity}(f, d) = \sum_{e \in \textit{edges}(f)} \sum_{e' \in h(e)} \textit{edge\_score}(e') 
\end{equation}

\paragraph{Fragment translation score.}
Concepts have a translation score, i.e., how well they represent the user's input (remember, users enter terms, not concept identifiers). 
Next, we define how well a fragment represents the user's intended information need.
A translated fragment close to the user's input is ranked higher.
The translation score for a fragment $f$ is defined as:
\begin{equation}
    \textit{translation}(f) = \min(\lbrace \textit{translation\_score}(c) \mid c \in \textit{nodes}(f) \rbrace)
\end{equation}

\paragraph{Weighting.} 
Some strategies come with scores between 0 and 1, while others, e.g., tf-idf, may yield scores above 1.0. 
Given a document set $D_r$ to rank, we normalize all scored fragments by their maximum score for each of our previous strategies. 
Let $D_r$ be the set of all documents to rank, $f$ be the matching fragment of document $d$ and $d$ the actual document to rank. 
We combine our four similarities $sim = [\textit{confidence}, \textit{min\_tfidf}, \textit{coverage}, \textit{relational\_similarity}]$ by weighting each one through a vector $W = [w_1, w_2, w_3, w_4]$ with $w_i \in [0, 1]$ and  $w_1 + w_2 + w_3 + w_4 = 1$.

\begin{equation}
   \textit{fscore}(f, d) = \textit{translation}(f, d) \cdot \sum_{w_i \in W} w_i \cdot sim_i(f, d)
\end{equation}

\paragraph{Document scoring.}
Each document graph might have multiple fragments that match the initial query. 
We compute each fragment's score. 
Here, we multiply the fragment's \textit{score} with its translation score. 
We then select the best-scored fragment to represent the overall document score for a document $d$:

\begin{equation}
    \textbf{GraphRank}(q, d) = \text{max}(\lbrace \textit{fscore}(f, d) \mid f \in \textit{matches}(q, d) \rbrace) 
\end{equation}

\subsection{Query Relaxation}
In the following, we introduce two paradigms to expand queries and thus, improve the system's recall.

\begin{figure}[t]
    \centering
    \includegraphics[width=\textwidth, trim={0cm 9.2cm 8cm 0cm}, clip]{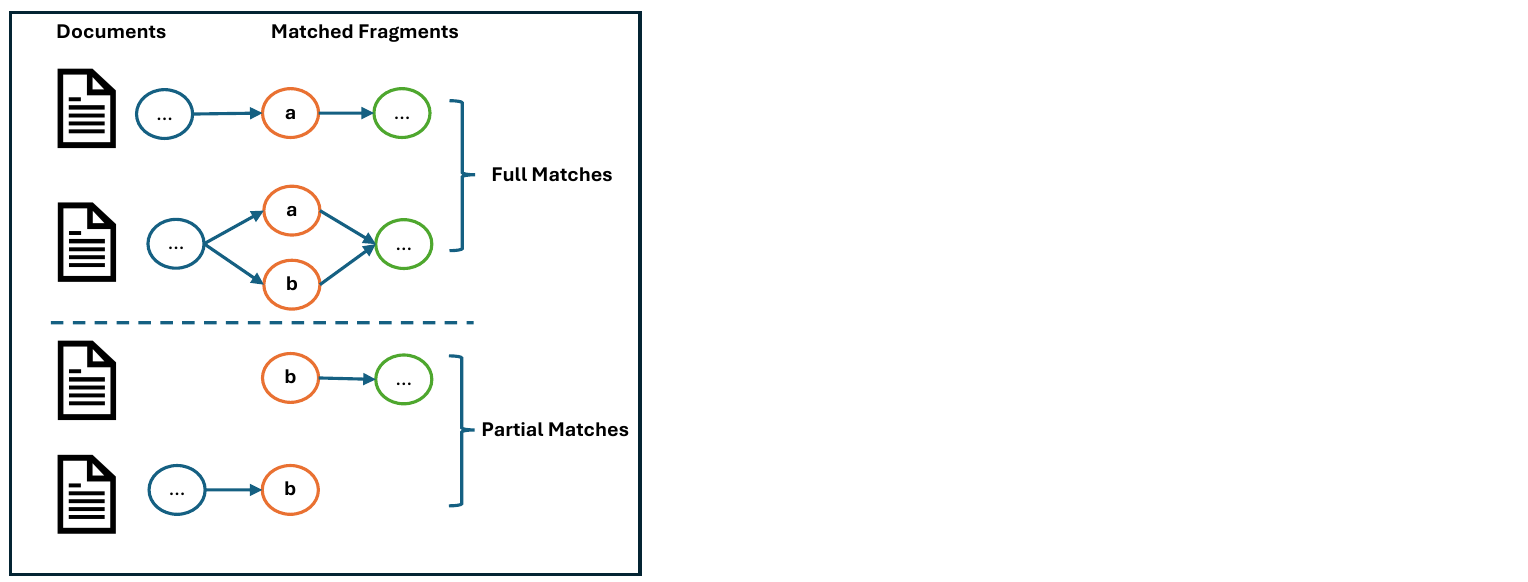}
    \caption{Conceptual Overview: When ranking document result lists with Full and Partial Matches, the list of Full Matches is always placed at the top of the final result list.}
    \Description{Conceptual Overview: When ranking document result lists with Full and Partial Matches, the list of Full Matches is always placed at the top of the final result list.}
    \label{fig:PartialMatches}
\end{figure}

\paragraph{Partial Matches.}
The retrieval system enforces that relevant documents must match the full graph query (\textbf{Full Match}). 
However, relaxing the query paradigm for more extensive and comprehensive result lists can be beneficial. 
Especially if the full match only returns a few documents, the document list could be expanded by documents that answer the query partially, i.e., that only contain a single fact pattern of the query (\textbf{Partial Match}).
We implement the Partial Match strategy as an extension for our discovery system. 
The Partial Match strategy enforces that documents that match the query fully are always placed before partial matches. 
In other words, the list of full document matches is ranked and the list of partial document matches is also ranked but is put behind the full matching document list, i.e., no partial document match can be better than a full document match.
Documents that match the query fully and partially are removed from the partial list, as they already matched the documents fully. 
A illustration is shown in Figure~\ref{fig:PartialMatches}.

\paragraph{Ontological Expansion.}
\label{sec:ontological_expansion}
Concepts in queries are by default expanded by their subclasses, e.g., if users search for general \textit{diabetes}, queries will also search for particular subtypes like \textit{diabetes type 2}. 
This decision was made when implementing our system~\cite{Kroll2023IJDL} because it reflected the users' needs. 
However, going upwards in an ontology might also be helpful; for example, consider more general forms of metabolic disease. 
A feature like that is yet missing because result lists can become very large.
Very large result lists, especially if they are not sorted by relevancy, can overload or confuse users in the end.
And a ranking paradigm was yet missing in the discovery system (it ranked documents by their publication date).

However, our proposed \textbf{GraphRank} approach allows us to rank those document lists with regard to their content relevancy concerning the query.
With that, users can expect precise hits in the top of the list while still having the option to explore hits further down in the list.
A concept $c$ can be generalized by the $\textit{superclass}(c)$ relation that retrieves all direct and transitive superclasses of $c$. 
Such a function can be implemented using ontologies like the Medical Subject Headings.
However, each step in the ontology we make might lead to more irrelevant results. 
That is why we introduce a similarity score for expanded concepts.
Let $a$ and $b$ be two concepts. 
We define their ontological similarity with regard to some ontology:

\begin{equation}
    \textit{ontological\_sim}(a, b) = \begin{cases}
 1 &, \textit{if } a = b \\
\frac{1}{|\textit{path}(a, b)|} &, \textit{if } \textit{path}(a, b) \neq \emptyset \\
 0 &, \textit{otherwise}\\
\end{cases}
\end{equation}

With that, we can expand a query upwards in an ontology like the following:
\begin{enumerate}
    \item[] For each fact pattern $fp$ in a query $q$:
    \begin{enumerate}
        \item Iterate over $fp's$ subjects and objects (concept sets)
        \item Apply the \textit{superclass} operator to each concept
        \item  Add each expanded concept to the corresponding list of subjects/objects
        \item Score this new concept by $ontological\_sim(c, super) \cdot \textit{translation\_score}$ of the source concept that has been expanded. 
    \end{enumerate}
\end{enumerate}

In this way, we rewrite the subject and object lists of each fact pattern and score the generalized concepts with regard to their distance within the used ontology. 
The more steps we take, the less \textit{well-translated} is the concept in reflecting the query.
\newpage

\section{Evaluation}
Our method GraphRank is designed to rank concept-centric narrative query graphs in the biomedical domain.
As far as we know, graph-based biomedical document retrieval benchmarks do not exist.
Crafting a comprehensive one is especially challenging due to the need for experts and relevance annotation. 
That is why we decided to focus on existing biomedical benchmarks that could be used for our purposes.
For instance, the TREC Precision Medicine Series 2017-2020~\cite{DBLP:conf/trec/RobertsDVBH20,DBLP:conf/trec/RobertsDVHBL18,DBLP:conf/trec/RobertsDVHBLP17,DBLP:conf/trec/RobertsDVHBLPM19} were designed as concept-centric document retrieval benchmarks.

\subsection{Evaluation Setup}
The central problem when utilizing these benchmarks is that they ask for keyword queries instead of graph queries. 
More precisely, no information about the relationships between keywords is given, which would be necessary to craft graph patterns. 
However, these queries usually ask for several biomedical concepts that interact with each other.
Consider, for example, the benchmark query \textit{melanoma BRAF Binimetinib} that asks for three biomedical components. 
We assumed the predicate was not given in the benchmark and allowed any predicate between the searched concepts.
With that assumption, we could generate a graph pattern like $(C_1, ?p_1, C_2) \wedge (C_2, ?p_2, C_3)$ which asks for some interaction between $C_1$ (all translated concepts for the first component) and $C_2$, and some interaction between $C_2$ and $C_3$.
A document then matches the query if it contains both interactions.
If a query asks for three components, we have three alternatives to connect the different components. 
Finally, we can generate the following graph query by using a logical disjunction over all three combinations: $[(C_1, ?p_1, C_2) \wedge (C_2, ?p_2, C_3)] \lor [(C_1, ?p_1, C_2) \wedge (C_1, ?p_2, C_3)] \lor [(C_1, ?p_1, C_3) \wedge (C_2, ?p_2, C_3)]$.
The number of alternatives depends on the number of components.

\paragraph{Restricted Concept Vocabulary.}
Please note that we did not adjust the system's concept vocabulary for this paper. 
Our subsequent evaluation reveals some major limitations here, i.e., queries or terms in queries that are not reflected in our system's concept vocabulary. 
However, extending the vocabulary was out of the scope of this work.
That is why we used the existing one and showed its advantages if queries can be successfully translated and its disadvantages if the query translation fails or the representation is not perfect.

\subsection{Benchmarks}
The \textbf{TREC-Precision Medicine} (PM) series (2017-2020)~\cite{DBLP:conf/trec/RobertsDVBH20,DBLP:conf/trec/RobertsDVHBL18,DBLP:conf/trec/RobertsDVHBLP17,DBLP:conf/trec/RobertsDVHBLPM19} consists of seven benchmarks for biomedical document retrieval. 
PM2020 was based on MEDLINE abstracts, whereas 2017-2019 had two evaluation settings: MEDLINE abstract retrieval and clinical trial retrieval. 
Clinical trials are different from abstracts written in natural language: 
They contain semi-structured enumerations and information. 
Our system's information extraction methods will yield a poor document representation because they rely on natural language texts.
That is why we focused on benchmarks with documents written in natural language. 
Unfortunately, the benchmarks in 2017-2019 ask for demographic information (e.g., 58-year-old men) and complex gene variations that are not supported by the retrieval system. 
The system's concept vocabulary only included \textit{basic} genes and did not include patient age information.
For instance, the system supports the search for the gene \textit{BRAF} but not for the variant \textit{(V600E)} in \textit{BRAF (V600E)}.
Similarly, modification of genes like \textit{Amplification} in \textit{CDK4 Amplification}, or \textit{Deletion} in \textit{CDKN2A Deletion} were not supported.
The PM2020 topics ask for drugs, diseases, and genes, which the system supports.
In other words: PM2020 evaluates well-supported queries, while PM2017-2019 have some drawbacks due to bad query translations.

We used \textbf{TREC-COVID 2020}~\cite{DBLP:journals/jbi/RobertsABDLSVWH21} as a fifth benchmark. 
TREC-COVID 2020 is a biomedical document retrieval benchmark that asks for topics about COVID-19.
One specialty is that the benchmark was crafted by asking \textit{common} people for their information needs.
In contrast to the TREC Precision Medicine Series, this benchmark comes with general topics like \textit{school closing coronavirus} and \textit{differences between flu and coronavirus}. 
We used this benchmark to show the limitation of the system: 
What happens if queries do not ask for precise biomedical concept interactions? 
For reporting purposes, we decided first to report PM2020 (well-supported scenario), move to PM2017-2019 (partially supported), and finally, TREC-COVID (not well-supported). 
With that we discuss advantages and drawbacks of overall approach.

\textbf{Data Processing.}
The retrieval system already contained all MEDLINE abstracts required to evaluate PM2017-2020. 
We downloaded each benchmark's MEDLINE articles, parsed them, and derived a list of included PubMed IDs. 
This list allows us to only query for documents included in one of the benchmarks (e.g., PM2017 asks for articles published before 2017 and so on).
PM2017-2019 include a list of additional biomedical abstracts not included in the MEDLINE.
The TREC-COVID benchmark comes with its own collection that includes abstracts and full texts. 
We separately evaluate both TREC-COVID settings, abstract and abstract + full-text retrieval. 
We used the same pipeline initially used to process the MEDLINE collection for the missing documents (additional abstracts and TREC-COVID documents). 
Scripts can be found in our repository.

\textbf{Query Translation.}
The benchmark asks for the following number of topics: 30 (PM2017), 50 (PM2018), 40 (PM2019), 31 (PM2020), and 50 (TREC-COVID). 
The PM Series splits query topics into components, such as disease, gene, treatment option, and patient information. 
We used that information for the query translation, e.g., a disease component is only translated into disease concepts. 
For our evaluation, we counted how well queries could be translated. 
Remember, a benchmark query asks for components like 1. disease, 2. gene and 3. drug. 
Therefore, we defined the query translation score as the minimum translation score of one of its components.
We used the maximum concept translation score to represent the translation score for a component.
The idea was that a query translation is as bad as its worst translatable component, and a component is as good as its best matching concept. 

In contrast to the PM series, TREC-COVID does not split topics into components. 
Instead, TREC-COVID offers one query string per topic, e.g., \textit{differences between flu and coronavirus}.  
We used a greedy search to map possible parts of each string to concepts (details are available at our GitHub): 
First, we checked whether the whole string can be mapped to a concept. 
If not, we remove the last token (split by space) and test whether the remaining string can be mapped to a concept. 
If not, we repeat the last step until we find a concept or end up with the most left token (here \textit{differences}). 
If the last remaining token (here \textit{differences}) cannot be mapped, we remove the first token and start with the first step again.
Here, we can identify \textit{flu} and \textit{coronavirus} as two, translatable token sequences.

\textbf{Ontological Expansion.}
Each TREC Precision Medicine Series topic asks for a specific type of cancer. 
The benchmark instructions state that the more precise a document's included cancer type corresponds to the searched one, the more relevant it is.
This benchmark instruction reflects our idea of ontological query rewriting. 
We used the Medical Subject Heading cancer ontology (see \url{https://meshb.nlm.nih.gov/record/ui?ui=D009369}) to rewrite our queries, i.e., we expand specific cancer types to general ones.

\subsection{Results}
\textbf{Parameters.}
For our evaluation, we used equal weights for our GraphRank method, i.e., $w_i = 0.25$.
We set predicate specificity score (see tf-idf score) based on each predicate's hierachical level in our three-level predicate taxonomy (most-specific predicates received a score of 1.0, one level higher 0.5, and the highest level (only associated) 0.25); see taxonomy at \url{https://narrative.pubpharm.de/help/}.
The idea is that the deeper a predicate is placed in our taxonomy, the more information a predicate carries.
Future work could perform some parameter search here to find better parameters.

\textbf{Baselines.} We compare GraphRank to the well-known ranking strategy BM25: 
First, we used BM25 to rerank the documents retrieved by the graph matching paradigm (BM25 Reranking). 
This setup compares GraphRank to BM25 on the same set of retrieved documents. 
Second, we used BM25 to retrieve and rank documents without graph-based retrieval. 
This setup demonstrates how graph-based retrieval plus ranking performs compared to pure BM25 retrieval (BM25 Native).
We used PyTerrier~\cite{DBLP:conf/ictir/MacdonaldT20} to implement BM25 and to index the documents.

\textbf{Evaluation.} During our evaluation, we observed that our graph-based matching retrieved many documents that had not been judged in the benchmarks. 
For instance, we retrieved 44 not-judged documents at rank 20 and 113 at rank 30 for PM2020.
A sample revealed that some of these documents contained synonyms or similar, related concepts of the searched components. 
We assume the benchmark authors used term-based strategies to collect the documents to judge. 
For our evaluation, we thus decided to ignore unjudged documents because our goal is to extend the discovery system and not to outperform other strategies.

\begin{table*}[t]
\caption{Evaluation results of TREC-PM2020~\cite{DBLP:conf/trec/RobertsDVBH20} (based on 31 out of 31 topics): Recall, nDCG@k and P@k are reported at different ranks. We show the results of the Old System, GraphRank in different combinations and BM25.}
\centering
\begin{tabular}{lccccccc}
\toprule
Ranking Method & Recall@1000 & nDCG@10 & nDCG@20 & nDCG@100 & P@10 & P@20 & P@100 \\
\midrule
Old System~\cite{Kroll2023IJDL} & 0.31 & 0.37 & 0.37 & 0.36 & 0.42 & 0.38 & 0.21 \\

\midrule  

\textbf{Full Match} & 0.31 & 0.37 & 0.37 & 0.36 & 0.42 & 0.38 & 0.22 \\
\quad + GraphRank & 0.31 & 0.42 & 0.41 & 0.38 & 0.45 & 0.40 & 0.22 \\
\quad + BM25 & 0.31 & 0.46 & 0.43 & 0.40 & 0.45 & 0.39 & 0.22 \\
\quad + Ontology & \textbf{0.45} & 0.33 & 0.33 & 0.37 & 0.41 & 0.36 & 0.24 \\
\quad + Ontology + GraphRank & \textbf{0.45} & 0.44 & 0.43 & 0.43 & \textbf{0.48} & \textbf{0.42} & \textbf{0.25} \\
\quad + Ontology + BM25 & \textbf{0.45} & \textbf{0.47} & \textbf{0.46} & \textbf{0.45} & 0.47 & \textbf{0.42} & \textbf{0.25} \\

\midrule  

\textbf{Partial Match} & 0.78 & 0.40 & 0.42 & 0.48 & 0.46 & 0.42 & 0.29 \\
\quad + GraphRank & 0.78 & 0.50 & 0.49 & 0.50 & \textbf{0.55} & \textbf{0.47} & 0.28 \\
\quad + BM25 & 0.78 & \textbf{0.53} & \textbf{0.52} & \textbf{0.55} & 0.53 & \textbf{0.47} & \textbf{0.30} \\
\quad + Ontology & \textbf{0.86} & 0.33 & 0.34 & 0.44 & 0.41 & 0.36 & 0.28 \\
\quad + Ontology + GraphRank & \textbf{0.86} & 0.47 & 0.48 & 0.51 & 0.52 & \textbf{0.47} & 0.29 \\
\quad + Ontology + BM25 & \textbf{0.86} & 0.50 & 0.51 & \textbf{0.55} & 0.51 & \textbf{0.47} & \textbf{0.30} \\
\midrule
\textbf{Native BM25 (Baseline)} & 0.79 & 0.48 & 0.46 & 0.49 & 0.48 & 0.42 & 0.28 \\
\bottomrule
\end{tabular}
\label{tab:pm2020}
\end{table*}

\begin{figure*}[ht]
    \centering
    \includegraphics[width=\textwidth, trim={4cm 0cm 4cm 1cm}, clip]{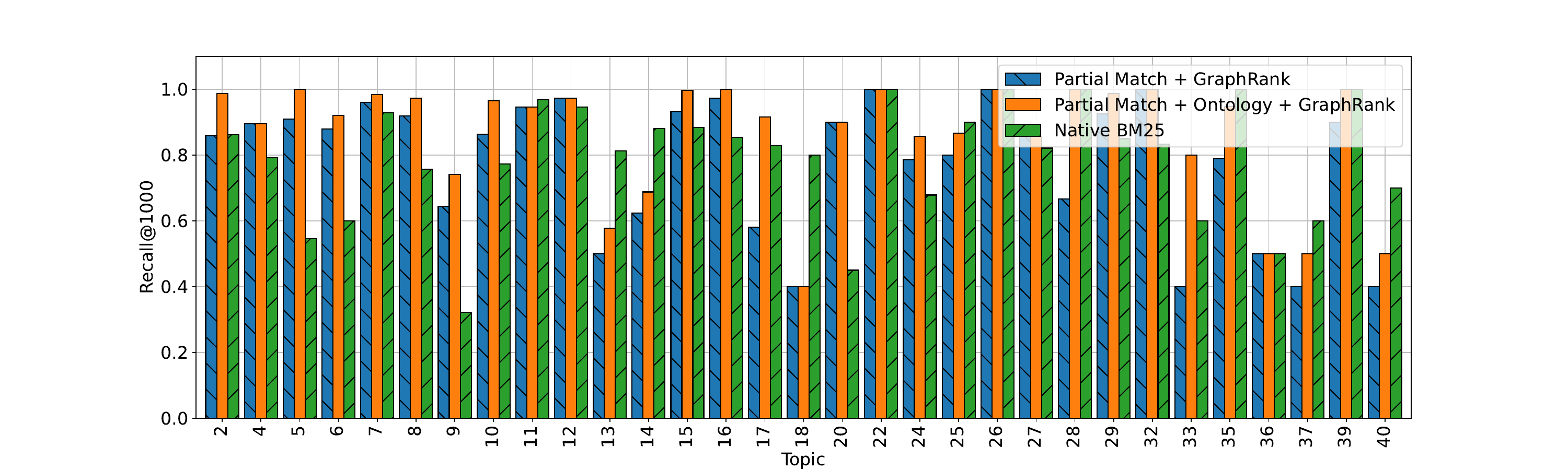}
    \caption{Topic-wise Recall@1000 evaluation on PM2020: GraphRank vs. Native BM25. }
    \Description{Topic-wise Recall@1000 evaluation on PM2020: GraphRank vs. Native BM25. }
    \label{fig:pm2020_ranker_topic_wise_like_recall}
\end{figure*}

\textbf{PM2020.}
For PM2020, 31 out of 31 queries had a translation score above 0.9. 
The results for PM2020 are depicted in Table~\ref{tab:pm2020} which has four parts: 1) the old system~\cite{Kroll2023IJDL} without the improved query translation and by sorting documents by their IDs (date) in descending order (old system), 2) using Full Match as a matching paradigm without ranking (just Full Match), plus ranking strategies (+GraphRank and +BM25) and ontological query expansion (+Ontology), 3) Partial Match without ranking, plus ranking and query expansion, and 4) the results for native BM25 retrieval. 
We report the Recall@1000, the normalized discounted cumulative gain (nDCG), and precision at different ranks k (@10, @20, @100). 
In summary, the recall of the Full Match paradigm is always below the Partial Match paradigm, which we expected. 
Partial Match plus ontological expansion achieved a recall of 0.86. 
In comparison, BM25 achieved a recall of 0.79. 
The type of queries can explain the high recall of BM25: 
Queries in PM2020 ask for a specific cancer subtype. 
Each subtype contained a term like \textit{cancer} (e.g., ovarian cancer). 
If the exact form cannot be found in documents, BM25 will also rank documents that \textit{just} contain the term \textit{cancer}.
In other words, BM25 performed some form of beneficial expansion here (compare it to our ontological rewriting). 
Figure~\ref{fig:pm2020_ranker_topic_wise_like_recall} shows a topic-wise Recall@1000 evaluation of PM2020. 
Partial Match + Ontology + GraphRank achieved a recall $\geq$ 0.9 in 19 out of 31 topics, whereas BM25 achieved nine times a recall $\geq$ 0.9.

\begin{figure*}[ht]
    \centering
    \includegraphics[width=\textwidth, trim={4cm 0cm 4cm 1cm}, clip]{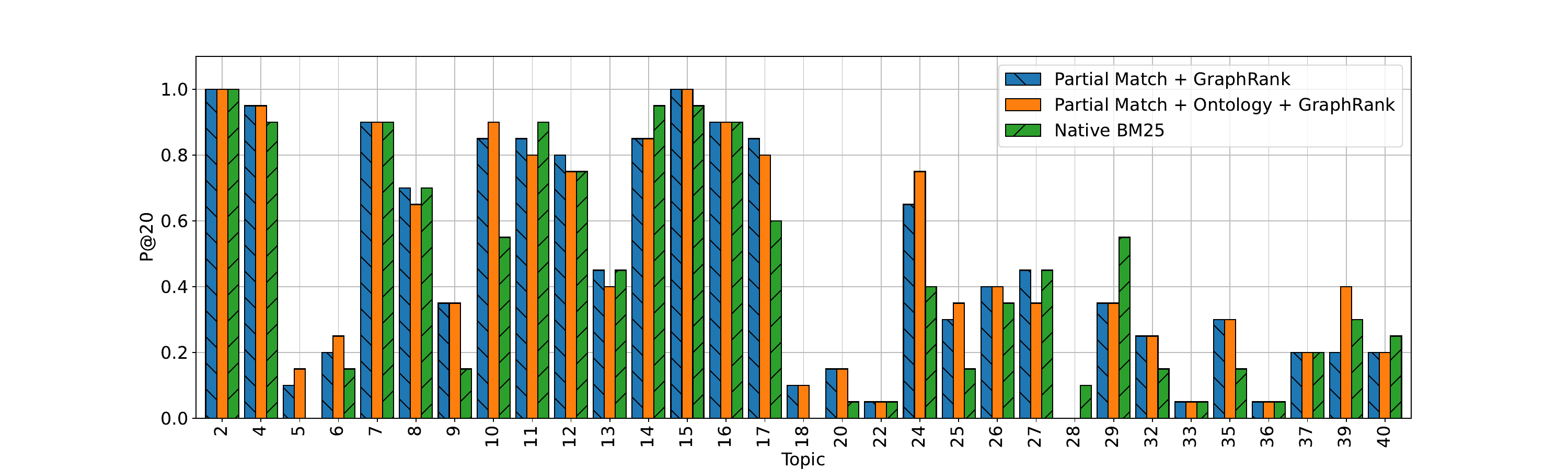}
    \caption{Topic-wise P@20 evaluation on PM2020: GraphRank vs. Native BM25}
    \Description{Topic-wise P@20 evaluation on PM2020: GraphRank vs. Native BM25}
    \label{fig:pm2020_ranker_topic_wise_like_p20}
\end{figure*}

\begin{figure*}[]
    \centering
    \includegraphics[width=\textwidth, trim={4cm 0cm 4cm 1cm}, clip]{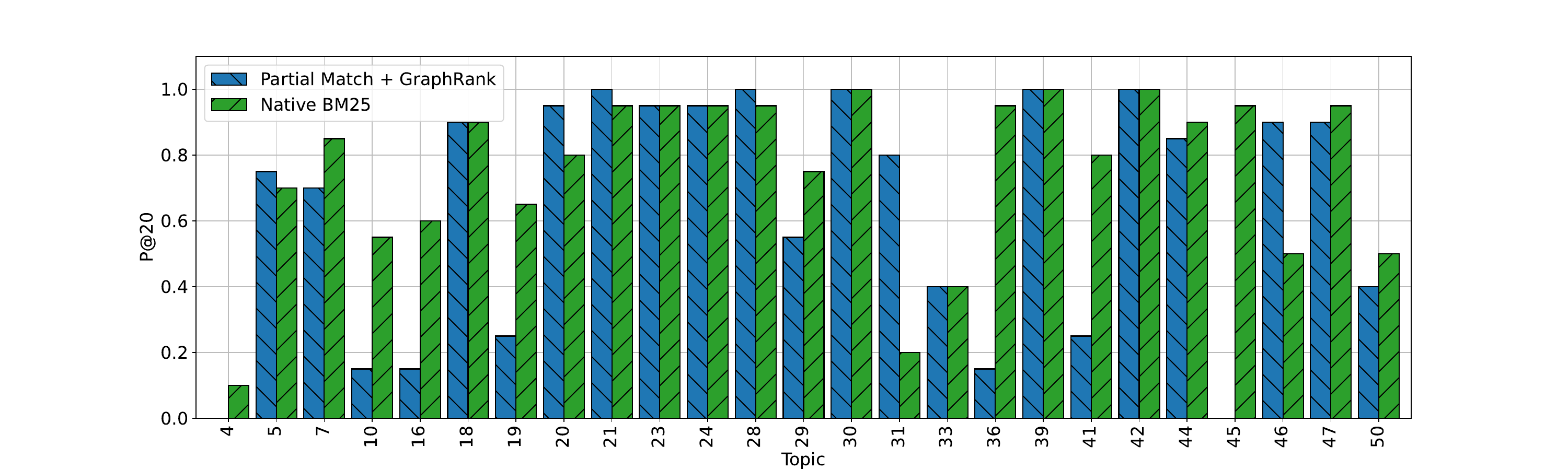}
    \caption{Topic-wise P@20 evaluation on TREC-COVID (Abstracts): GraphRank vs. Native BM25}
    \Description{Topic-wise P@20 evaluation on TREC-COVID (Abstracts): GraphRank vs. Native BM25}
    \label{fig:trec_covid_ranker_topic_wise_like_p20}
\end{figure*}

Concerning precision, Partial Match plus BM25 or GraphRank achieved higher scores than native BM25 retrieval (up to 7\% points in P@10).  
Partial Match did not decrease the precision in comparison to Full Match.
In contrast, it increased the precision because:
First, Partial Match puts partially matched documents behind full matches in a result list (by definition).
Second, for nine topics, Full Match yielded less than 20 results, decreasing the precision at rank 20.
For instance, if five correct matches were found but nothing more, the precision at 20 is 0.25 by definition. 
Compared to performing a BM25-reranking of documents retrieved by Partial Match and Ontology, GraphRank achieved a comparable performance (slightly better/comparable for precision, but slightly worse for nDCG).
Figure~\ref{fig:pm2020_ranker_topic_wise_like_p20} shows a topic-wise evaluation concerning P@20.
Further figures, e.g., with different metrics and k, as well as an ablation study are available in our repository.

\begin{table*}[ht!]
\caption{Evaluation on PM2017~\cite{DBLP:conf/trec/RobertsDVHBLP17}, PM2018~\cite{DBLP:conf/trec/RobertsDVHBL18}, PM2019~\cite{DBLP:conf/trec/RobertsDVHBLPM19} and TREC-COVID~\cite{DBLP:journals/jbi/RobertsABDLSVWH21}. Recall, nDCG@k and P@k are reported at different ranks. We show the results of GraphRank in different combinations and BM25.}
\centering
\begin{tabular}{lccccc}
\toprule
Strategy & Recall@1000 & nDCG@10 & nDCG@20 & P@10 & P@20\\
\midrule
\multicolumn{6}{c}{\textbf{TREC-PM 2017 - Topics (26/30) - Abstracts}} \\
\midrule
Partial Match & 0.6 & 0.38 & 0.37 & 0.43 & 0.40 \\
\quad + GraphRank & 0.6 & \textbf{0.47} & \textbf{0.46} & \textbf{0.5} & \textbf{0.48} \\
\quad + Ontology + GraphRank & \textbf{0.84} & 0.42 & 0.42 & 0.45 & 0.46 \\
Native BM25 (Baseline) & 0.56 & 0.41 & 0.39 & 0.41 & 0.39 \\
\midrule
\multicolumn{6}{c}{\textbf{TREC-PM 2018 - Topics (45/50) - Abstracts}} \\
\midrule
Partial Match & 0.76 & 0.44 & 0.44 & 0.51 & 0.49 \\
\quad + GraphRank & 0.76 & \textbf{0.53} & \textbf{0.54} & \textbf{0.62} & \textbf{0.62} \\
\quad + Ontology + GraphRank & \textbf{0.87} & 0.44 & 0.48 & 0.54 & 0.57 \\
Native BM25 (Baseline) & 0.64 & 0.51 & 0.50 & 0.52 & 0.51 \\
\midrule
\multicolumn{6}{c}{\textbf{TREC-PM 2019 - Topics (37/40) - Abstracts}} \\
\midrule
Partial Match & 0.62 & 0.42 & 0.43 & 0.48 & 0.47 \\
\quad + GraphRank & 0.62 & \textbf{0.49} & \textbf{0.50} & \textbf{0.54} & \textbf{0.54} \\
\quad + Ontology + GraphRank & \textbf{0.82} & 0.40 & 0.43 & 0.46 & 0.49 \\
Native BM25 (Baseline) & 0.63 & 0.48 & 0.47 & 0.51 & 0.48 \\
\midrule
\multicolumn{6}{c}{\textbf{TREC COVID 2020 - Topics (25/50) - Abstracts}} \\
\midrule
Partial Match & 0.26 & 0.54 & 0.51 & 0.61 & 0.57 \\
\quad + GraphRank & 0.26 & 0.61 & 0.59 & 0.68 & 0.64 \\
Native BM25 (Baseline) & \textbf{0.47} & \textbf{0.71} & \textbf{0.68} & \textbf{0.79} & \textbf{0.75} \\
\midrule
\multicolumn{6}{c}{\textbf{TREC COVID 2020 - Topics (25/50) - Fulltexts}} \\
\midrule
Partial Match & 0.39 & 0.52 & 0.49 & 0.58 & 0.54 \\
\quad + GraphRank & 0.39 & 0.58 & 0.56 & 0.64 & 0.63 \\
Native BM25 (Baseline) & \textbf{0.48} & \textbf{0.71} & \textbf{0.70} & \textbf{0.80} & \textbf{0.78} \\
\bottomrule
\end{tabular}
\label{tab:pm2017_2018_2019}
\end{table*}

\textbf{PM2017-2029.}
The PM2017, 2018, and 2019 results can be found in Table~\ref{tab:pm2017_2018_2019}. 
26 out of 30 (PM2017), 45 out of 50 (PM2018), and 37 out of 40 topics (PM2019) could be translated into a graph representation with a translation score above 0.9.
A manual evaluation revealed that the remaining topics asked for specific diseases or genes not included in the system's concept vocabulary. 
In summary, GraphRank always boosted precision and nDCG compared to Partial Match without ranking.
Partial Match + GraphRank also received higher scores (recall, precision, and nDCG) than Native BM25 retrieval.
Adding an ontological expansion again (+ Ontology) clearly improved the recall, e.g., from 0.6 to 0.84 on PM2017, 0.76 to 0.87 on PM2018, and 0.62 to 0.82 on PM2019.

\textbf{TREC-COVID.}
24 out of 50 topics had a translation score above 0.9 and contained at least two concepts.
On TREC-COVID, native BM25 retrieval outperformed Partial Match + GraphRank in both settings (abstract/full-text retrieval) and all metrics; see Table~\ref{tab:pm2017_2018_2019}.  
A detailed topic-wise evaluation, see Figure~\ref{fig:trec_covid_ranker_topic_wise_like_p20} reveals that GraphRank achieved high scores ($\ge$ 0.9) in eleven topics that ask for biomedical concepts (e.g., T21 - \textit{coronavirus mortality}, T23 - \textit{coronavirus hypertension},
T39 - \textit{COVID-19 cytokine storm}). 
In topics like T4 - \textit{how do people die from the coronavirus} or T45 - \textit{coronavirus mental health impact}, our approach might know some concepts (e.g., \textit{coronavirus}, \textit{people} or \textit{impact}, otherwise the query would not be translatable), but the matched concepts do not reflect the full information need. 
In summary, concept-centric queries were answered well.
However, 25 out of 50 topics could not be translated at all, and some translated topics were answered poorly because the information needs were not well reflected due to missing concepts or not represented terms.

\subsection{Discussion}
While PM2017-2020 verified that many topics could be translated successfully into graph patterns, TREC-COVID showed clear limitations here, as only about half of the topics could be represented as graphs.
However, TREC-COVID topics are very general, e.g., \textit{school closings during coronavirus}, whereas the graph-based retrieval system was designed to answer scientific, concept-centric biomedical information needs instead.
Indeed, we argue that precise, concept-centric information needs are common in the biomedical domain, see the TREC Precision Medicine Series, an user study in~\cite{Kroll2023IJDL}, the graph-based system's query log analysis in~\cite{DBLP:conf/jcdl/KrollKSB23}, or this PubMed query log analysis~\cite{herskovic2007pubmedqueryanalysis}.
This paper extends our existing graph-based retrieval system~\cite{Kroll2023IJDL} by adding a ranking strategy (GraphRank), a relaxed query paradigm (Partial Match), and ontological query rewriting. 
Moreover, we demonstrated that these methods effectively answered biomedical information needs in terms of precision through GraphRank and recall through Partial Match and ontological rewriting.
While we achieved a comparable, or sometimes slightly improved, performance compared to re-ranking graph matches with BM25, GraphRank allows us to work solely on the graph structure of documents.
From a system perspective, the main advantage is that we could directly integrate our methods into the existing discovery system.
In other words, we do not need to build, maintain, and update a second index based on textual features, e.g., to calculate BM25.
Instead, we can directly compute the graph-based scores on-the-fly when retrieving the documents.

\section{Conclusion}
Benefits of graph-based retrieval, like entity-centric structured overviews of the literature have already been discussed in the literature~\cite{kroll2022nia,Kroll2023IJDL}.
However, practical ranking methods for such retrieval systems, solely based on the graph-based document representation, were yet missing.
In this work, we filled that gap by proposing methods for such retrieval workflows, which opens up a space for future research. 
Therefore, we share our implementation as an example implementation for graph-based discovery and ranking workflows.
Moreover, we proposed effective query relaxation and ontological rewriting that can improve recall and thus help users explore a document collection. 
Our methods work on an extensive digital library collection with 37M documents, do not require training data or supervision, and can be directly integrated into an existing digital library system.
If queries were concept-centric and the system knew those concepts, our methods outperformed BM25.
However, not all information needs could be translated successfully because concepts were missing (school closing) or lack of expressiveness (gene modifications). 
Future work could tackle a fallback mode for switching between graph-based and traditional text-based ranking, depending on a certain information need.

\section*{Acknowledgments}
Supported by the Deutsche Forschungsgemeinschaft (DFG, German Research Foundation): PubPharm – the Specialized Information Service for Pharmacy (Gepris 267140244).

\bibliographystyle{ACM-Reference-Format}
\bibliography{references}

\end{document}